# Accelerated chemical space search using a quantum-inspired cluster expansion approach

Hitarth Choubisa*,[1], Jehad Abed*,[1], Douglas Mendoza[2,3], Hidetoshi Matsumura[4], Masahiko Sugimura[4], Zhenpeng Yao[2,3,6], Ziyun Wang[1,5], Brandon Sutherland[1], Alán Aspuru-Guzik[6,7,8,9,†], Edward H Sargent[1,†]

[1]Department of Electrical and Computer Engineering, University of Toronto, 10 King's College Road, Toronto, ON, M5S 3G4, Canada.

[2]Department of Chemistry, University of Toronto, 80 St. George Street, Toronto, ON M5S 3H6, Canada

[3]Department of Chemistry and Chemical Biology, Harvard University, 12 Oxford Street, Cambridge, MA 02138, USA

[4]Fujitsu Consulting (Canada) Inc, Toronto, ON, Canada

[5]School of Chemical Sciences, The University of Auckland, Auckland, 1010, New Zealand

[6]Chemical Physics Theory Group, Department of Chemistry, University of Toronto, Toronto, ON M5S 3H6, Canada

[7]Department of Computer Science, University of Toronto, Toronto, ON M5S 3H6, Canada

[8]Vector Institute for Artificial Intelligence, Toronto, ON M5S 1M1, Canada

[9]Lebovic Fellow, Canadian Institute for Advanced Research, Toronto ON, M5S 1M1, Canada

(*) These authors contributed equally to this work.

(†) Correspondence and requests for materials should be addressed to Alán Aspuru-Guzik (alan@aspuru.com) and Edward H. Sargent (ted.sargent@utoronto.ca)

**Abstract:**

To enable the accelerated discovery of materials with desirable properties, it is critical to develop accurate and efficient search algorithms. Quantum annealers and similar quantum-inspired optimizers have the potential to provide accelerated computation for certain combinatorial optimization challenges. However, they have not been exploited for materials discovery due to absence of compatible optimization mapping methods. Here we show that by combining cluster expansion with a quantum-inspired superposition technique, we can lever quantum annealers in chemical space exploration for the first time. This approach enables us to accelerate the search of materials with desirable properties order 10-50 times faster than genetic algorithms and bayesian optimizations, with a significant improvement in ground state prediction accuracy. Levering this, we search chemical space for discovery of acidic oxygen evolution reaction (OER) catalysts and find a promising previously unexplored chemical family of Ru-Cr-Mn-Sb-$O_2$. The best catalyst in this chemical family show a mass activity 8 times higher than state-of-art $RuO_2$ and maintain performance for 180 hours while operating at 10mA/$cm^2$ in acidic 0.5 M $H_2SO_4$ electrolyte.

**Introduction:**

Finding materials with desirable properties such as high d-band centers, high stability, high mobilities, optimal bandgaps, or low overpotentials to enable efficient energy harvesting[1], catalysis[2], light emission[3], sensing[4], and more is a challenging high-order optimization problem. Traditionally, density functional theory (DFT) based screening[5,6] has been used to explore chemical space. However, the high computational cost associated with DFT calculations[7] and the vastness of chemical space[8] make exhaustive searches infeasible. To predict material properties rapidly, studies in recent years have employed machine learning (ML) based surrogate models[9–13]. Among them, approaches based on learning from stoichiometries[14,15] and generative models (GMs) such as variational autoencoders (VAEs)[7,16–18] and generative adversarial networks (GANs)[19–22] allow accurate property prediction without DFT relaxation which is a computationally expensive step. However, despite the presence of an accurate ML model, chemical exploration can still be limited due to the generation of infeasible structures by generative models, absence of efficient search strategies and large chemical space (**SI note 1** for an estimate). Therefore, ensuring global optimization across an exhaustive search within chemical space is difficult with such methods. This is further complicated by the observation that formulating the chemical space search as an optimization problem is itself challenging due to the absence of a simple analytic optimization expression and the non-convex nature of structure-property relationships due to the presence of multiple local minima[23,24].

We aim to address this challenge by developing a new method to search chemical space by reframing the problem as one of finding the Ising Hamiltonian's ground state (**Figure 1**). Our method maps the Cluster Expansion (CE)[25,26] formulation to an Ising model. This mapping enables potential use of quantum annealers and QUBO solvers to perform rapid global searches in chemical space. We choose a quantum-inspired optimization engine (Digital Annealer[27]: henceforth referred to as DA; refer to Methods subsection on *DA* for more details) as the choice to solve for the ground state of our Ising model. Our choice is inspired by the fact that DA has been shown to efficiently find ground states of Ising models[28,29]. We term the approach **QCE** (**Quantum-inspired Cluster Expansion**). When used on its own, QCE allows us to circumvent the DFT relaxation process and simultaneously enable us to accelerate the search for materials that optimize target properties of interest (**Figure 1**). At the same time, QCE can also be used in conjunction with superior property prediction models such as interatomic potentials[10,30,31] that can accurately predict the properties to exhaustively search the material space and find materials with desirable properties.

In this paper we first detail the QCE method and benchmark it by finding stable materials in the quaternary Cu-Ni-Pd-Ag chemical system. Following the benchmark, we lever the framework to find stable and efficient acidic OER catalysts by developing an experimentally verified quantum mechanical catalyst efficiency proxy. Finally, we verify our predictions experimentally and our top catalyst (a never before reported multi-metal oxide) demonstrates best-in-class stability among all acidic rutile OER catalysts with overpotentials < 250 mV.

**Mapping cluster expansion to Ising Hamiltonian instance:**
Cluster expansions (CEs) are power series expansions of the partition function that account for many-body interactions. They have been widely employed for the exploration of materials[26,32–36]. In CE, the desired property P is expressed as a function of the 3D arrangement of atoms in the lattice. The 3D arrangement of atoms can be mapped bijectively to a configuration vector $\sigma = [\sigma_1, \sigma_2, \ldots, \sigma_N]$ where $\sigma_i$ represents the type of atom at a specific lattice site $i$. $N$ is the total number of sites in the lattice. The property of interest P is expressed as a function of the vector $\sigma$:

$$P(\sigma) = \sum_f J_f \Pi_f(\sigma) \tag{1}$$

where $J_f$ is the fitted correlation coefficient and $\Pi_f(\sigma)$ is the correlation function of multi-body interaction terms $\sigma$. By expanding the correlation functions in equation 1, we get:

$$P(\sigma) = \sum_{i \in cluster\_1} J_f \langle \Phi_i(\sigma_i) \rangle_f + \sum_{i,j \in cluster\_2} J_f \langle \Phi_i(\sigma_i) \cdot \Phi_j(\sigma_j) \rangle_f \tag{2}$$
$$+ \sum_{i,j,k \in cluster\_3} J_f \langle \Phi_i(\sigma_i) \cdot \Phi_j(\sigma_j) \cdot \Phi_k(\sigma_k) \rangle_f + \cdots$$

where $\Phi_i(\sigma_i)$ is an orthogonal basis function dependent on the atom type $\sigma_i$ [35]. Terms *cluster_1*, *cluster_2* and *cluster_3* are single-body, 2-body, and 3-body interactions of a unique cluster of sites. The $J_f$ coefficients are obtained by fitting the CE on the DFT-generated data (initial structures and relaxed structure properties)[36–38] (**Figure S1B**). The cluster expansion is an infinite series, but it can be approximated as a finite series[38,39].

The key step of our proposed mapping is expressing the basis functions $\phi_i(\sigma_i)$ as a superposition of all basis functions for every site within the lattice:

$$\Phi_i(\sigma_i) = \sum_{j=1}^{n} c(M_j) \cdot \phi_i(\sigma_i = M_j) \tag{3}$$
$$\sum_{j=1}^{n} c(M_j) = 1$$
$$c(M_j) \in \{0,1\} \; \forall \; j$$

such that $c(M_j)$ is the linear combination coefficient where the summation of the coefficients is equal to 1, $M_j$ is one chemical element from the periodic table and $j$ represents the index of that element ($j = 1 \ldots n$). It is due to the use of this superposition with constraints on coefficients that we name our method quantum-inspired cluster expansion (QCE).

In this study, we use the encoding: $c(M_j) = q_j$ where $q_j$ is a binary variable only accepting 0 and 1 as the values. Such variables are quite often realized as spins of a quantum annealer or binary decision variables in quantum-inspired optimizers. Several other candidate functions can also be used for the encoding scheme $c(\cdot)$ (refer to Methods subsection *encoding schemes for QCE* for an in-depth analysis of alternative encoding schemes). Using the encoding $c(M_j) = q_j$ transforms (3) as,

$$\Phi_i(\sigma_i) = \sum_{j=1}^{n} q_j \cdot \phi_i(\sigma_i = M_j) \tag{4}$$

Substituting for $\Phi_i$ from (4) in (2) yields an expression with several multi-body interaction terms ($q_0q_1$, $q_0q_1q_2$, $q_0q_1q_3q_4$,...). We use term reduction as reported by Babbush et al. [40] to simplify the expression repeatedly until a Quadratic Unconstrained Binary Optimization (QUBO) expression is obtained:

$$P = \sum_{i,j=1}^{N'} A_{ij} q_i q_j \tag{5}$$

where $A_{ij}$ are QUBO coefficients that are a function of $J_f$, $\Pi_f$ and penalties used for term reduction. $q_i$ is a binary decision variable accepting only 0 or 1 as the value. $N'$ is the total number of decision variables including the auxiliary variables used for term reduction.

The QUBO expression is then transferred to DA for efficient search and finding configurations that have the optimal property (minimum or maximum). Since any QUBO expression can be mapped to an Ising Hamiltonian, our proposed approach can be used by any Ising model solver such as a quantum annealer for optimizing materials within chemical space. This further motivated the term "quantum-inspired cluster expansion". The complete process is summarized in **Figure 1**.

### Benchmarking our approach through exploration of quaternary chemical space (Cu-Ni-Ag-Pd)

To benchmark the performance of our proposed approach, we explore the quaternary chemical space spanned by Cu-Ni-Ag-Pd in an FCC lattice due to their potential applications as fuel cell membranes[41] and oxygen reduction electrocatalysts[42]. Stability of materials for such applications is critical and is being extensively investigated[41,43]. Thus, finding stable alloys within this chemical space can be an exciting prospect[41,44] (**Figure 2a**). We use DFT for data generation and perform structure relaxation on randomly decorated FCC structures, followed by evaluation of the total energy of the systems. These total energies are then used to calculate mixing energies ($\Delta H_{mix}$) that are used to train a CE model (refer to Methods subsection *Data generation and mixing energy* for more details) and use it a measure of alloy's stability against its single metal constituents. The model achieves a 10-fold cross-validation MAE of 8 meV/atom (**Figure 2b;** methods subsection on cluster expansion training for more details). We used cluster expansion with 2-body and 3-body interactions with cut-off radius of 12Å as we found it to be most accurate in terms of cross-validation mean absolute error.

Using the scheme presented in the previous section, we prepared the QUBO representation of our CE model and transferred it to the DA for finding structures with minimum mixing energy (refer to methods subsection on *DA* for more details). To explore the potential of our approach, we sought to determine stable alloys within the Cu-Ni-Pd chemical subspace. We achieved this by restricting the possible values of the basis function $\Phi_i$ to only include relevant elements in the expansion of (3) (refers to methods subsection *Chemical subspace search* for more details). Upon performing the search using DA, we obtained equimolar ordered **Cu$_{0.5}$Pd$_{0.5}$** alloy as the alloy with most negative mixing energy ($\Delta H_{mix}$= -0.20 eV/atom) in Cu-Ni-Pd space and therefore, stable (**Figure S2**). This observation is independently supported by a previous report about equimolar FCC Cu-Pd alloy being the more stable[44].

Furthermore, we compared the temporal and efficiency performance of our approach to other widely used search algorithms in chemistry and material science: genetic algorithms and bayesian optimization. We implement the genetic algorithms (GA) through DEAP framework[45] and bayesian optimization (BO) using scikit-optimize[46] (for exact implementation, please refer to our code). We find that our approach when used in combination with the DA significantly outperforms the alternatives in finding the minimum mixing energy structure (**Figure 2c**; logarithmic scale). We also observe that our approach results in better quality of global minima. Even though the energy difference between QCE based predictions and best GA based predictions is 0.03 eV/atom, the ground state structures found by them are significantly different (Cu$_{0.5}$Pd$_{0.5}$ and Cu$_{0.69}$Pd$_{0.31}$ respectively) indicating the importance of accurate search algorithms. In addition, when we compare the results of DA against BO, we observe much worse performance than DA as well as GA. The final compositions found by BO were much closer to the starting points indicating that BO used in our comparisons is more effective in searching for local optimum than global optimum. All the tests were performed using a 20 core Intel Xeon(R) 6148 CPU operating at 2.4 GHz. Detailed parameters for all the comparison algorithms are presented in the Methods subsection *Performance benchmark algorithms*.

**Finding stable and efficient OER materials**

To demonstrate that the generalizability of our approach, we showcase the discovery of acidic oxygen evolution reaction (OER) catalysts[47,48]. The development of OER catalysts has been the subject of significant research over the past few years[43,49–54]. For the alkaline medium, several stable and efficient OER catalysts have been proposed; however, only a few stable, efficient, and cost-effective OER catalysts have been reported in literature for operation in an acidic medium[50,55].

High efficiency and high stability are two crucial requirements for a good OER catalyst. To develop a quantitative model for efficiency of the catalyst candidates, we built upon the d-band model by Hammer and Nørskov[56–59] and recent studies about its interplay with p-band centers[60]. We used materials and alloys constructed by substituting two different prototype structures: **ZrO$_2$** (P2$_1$/c; monoclinic) and **RuO$_2$** (P4$_2$/mnm; tetragonal) in our study. We generate slabs to model the catalytic activities for randomly generated structures(refer to Data Generation in Methods). After performing DFT calculations, we use the adsorption energies to determine the theoretical overpotentials (refer to **Figure S4** for model of the slabs). Performing logistic regression on the differential improvement of overpotential i.e., if overpotential of one material is smaller than the other (refer to logistic regression analysis in Methods; **Figure 3a**). We observe that probability of material $i$ having an overpotential smaller than materials $j$ is given by the following expression,

$$Prob(\eta_i < \eta_j) = \frac{1}{1+e^\theta} \tag{6}$$

$$where \;\; \theta = 0.078 \cdot (E_{p_i} - E_{p_j}) - 0.165 \cdot (E_{d_i} - E_{d_j})$$

where $\eta$ represent overpotential, $E_p$ and $E_d$ represent p-band center and d-band center respectively.

To further establish the validity of this theoretical analysis, we further conduct several in-lab experiments. Using just the contributions due to band centers, we achieve a Pearson's correlation coefficient of 0.67 (p-value < 0.001) between experimental overpotentials and theoretical proxy. We also observe that when combined with enthalpy of mixing ($\Delta H_{mix}$ per atom), we observe a significant improvement in predictive power using simple Ridge regression. Inclusion of mixing energy as a predictor significantly improves 10-fold cross-validation correlation from **0.67 to 0.81** (Figure 3b,c and **SI note 3** for more details). This role of mixing energy as a predictor can be explained as stable catalysts are likely to catalyze the reaction consistently. The overall expression to predict overpotentials $\eta$ is expressed as,

$$\eta_i = K(0.046 E_p - 0.097 E_d - \Delta H_{mix} + C) \tag{7}$$

where $E_p$ is p-band center, $E_d$ is the d-band center, $\Delta H_{mix}$ is mixing energy in eV/atom, $C$ is a constant and $K$ is just a multiplier. The values for the three properties are not on the same scale and that explains the large variation in the absolute value of coefficients. For exact values of $K$ and $C$, please refer to **SI note 3**. We also report all the collected experimental quantities for the candidates with measurable activity and acidic stability in **SI table 3** as a tabular database for future machine learning explorations and for reference. In addition, we also used mixing energy $\Delta H_{mix}$ to quantify stability of an alloy relative to its precursors i.e., ease of synthesizability of the alloy.

Following the strategy for QCE, we trained CE models (**Figure S5;** methods subsection on *cluster expansion training*) and generated the QUBO representations for each of the three properties: $\Delta H_{mix}(H_E)$, p-band center ($H_p$) and d-band center ($H_d$) using QCE. These properties were obtained by first performing DFT geometry optimization on the initial decorated lattices followed by calculation of the converged electronic and thermodynamic properties. Performing cross-validation analysis to find the optimal cut-offs, we find that CE models with 4Å and 8Å cut-offs for 2-body and 3-body interaction terms lead to best predictive power. Similarly, for predicting d-band centers, highest predictive power is demonstrated by a CE with 12Å and 6Å cut-offs for 2-body and 3-body interactions. CE model that incorporated just 2-body interactions with a cut-off of 4Å achieved the best predictive power. The accuracy and performance of each of these models on validation data are shown in **Figures 3d, 3e, and 3f**. We defined QUBO for efficiency as,

$$H_{eff} = 0.046 E_p - 0.097 E_d - \Delta H_{mix} \tag{8}$$

Thus, the problem of finding an efficient and stable OER catalyst transforms into finding a state that minimizes both $H_{eff}$ and $H_{stable}$ where $H_{stable}$ is the QUBO representation for mixing energies. We solve this problem by using a heuristic approach of optimizing the linear combination of the two QUBOs:

$$H_{mixed} = \lambda_1 H_{stable} + \lambda_2 H_{eff} \tag{9}$$

where $\lambda_1, \lambda_2 \in [0, \infty)$ (Refer to **Figure S6** for the effect of parameters $\lambda_1, \lambda_2$). Transferring $H_{mixed}$ represented as QUBO to DA and performing the optimization result in candidates that have promising stability as well as efficiency.

We search for different chemical subspaces within the two prototype structures by varying the relative weight parameters $\lambda_1, \lambda_2$ and chemical subspace of interest (**Figure S7** for details) while adding constraints to ratios of different elements (Refer to **figure 4** for a summary of candidates experimentally tested and **table S1** for a detailed list) whereas non-zero pair lead to a balanced trade-off between stability and efficiency **(Figure S6)**. We also compare this approach to a sequential search (search for stability followed by efficiency). We find that our heuristic approach outperforms a sequential search strategy (**SI note 7**). In our searches, we focus on exploring ternary and quaternary chemical spaces. This choice was motivated by existing exploration of binary transition metal oxides in literature[43,51] as well as effectiveness of solgel method of synthesis for multi-metal oxide synthesis.

In the context of the presented QCE method for chemical space search, it is also worth noting the previous works by Pedersen et al[61] on usage of Bayesian optimization for catalyst search. Both the observations made by the authors of that study and our benchmarking experiments indicate that a Bayesian optimization based search strategy is better suited to a local optimum search than a global optimum. So, if the question is to find a globally optimal material, QCE will outperform Bayesian optimization based approaches. On the other hand, both Bayesian optimization as well as QCE with constraints can be used for local optimizations of materials within the chemical space. With QCE, we can constraint the elemental fractions as inequalities and perform local chemical space search.

**Experimental verification:**

All of the 49 electrocatalyst candidates in **Table S1** were synthesized using sol-gel synthesis followed by annealing in air for 2 hours[62]. We observed that some compositions (8 compounds) were not successfully synthesized using the sol-gel method due to poor promotion of networks in the gel (**Figure 4a**). The crystal structure of the synthesized samples was screened using X-ray diffraction (XRD). XRD patterns were collected and classified into two groups: pure rutile phase (21 compounds) and multiple phases groups (20 compounds). One XRD pattern is shown in **Figure 4e** for $Ru_{0.58}Cr_{0.25}Mn_{0.08}Sb_{0.08}O_2$ (R14-M10-46) showing characteristic peaks of a rutile structure similar to a baseline $RuO_2$ catalyst. The XRD peaks are broad indicating that our synthesis approach produced nanocrystalline electrocatalysts. Further analysis of the structure for all electrocatalysts is provided in **Table S2**. High-resolution transmission electron microscopy (HR-TEM) and scanning transmission electron microscopy (STEM) are shown for three samples in **Figure S8**. The nanoparticles have an average particle size of *ca.* 10 nm with a spherical shape suitable for electrocatalytic applications. The energy X-ray dispersive spectroscopy (EDS) elemental mapping shows a homogeneous distribution of elements in the three samples (**Figure S9**). The EDS elemental mapping of $Ru_{0.58}Cr_{0.25}Mn_{0.08}Sb_{0.08}O_2$ (R14-M10-46) matches within 5% error with the nominal composition of the compound (Ru: 59±5.9 at. %, Cr: 20±1.0 at. %, Mn: 9±0.6 at. %, Sb 13±3.4 at. % of the total metal amount). The electron diffraction rings of the three samples confirm the formation of a strained rutile structure (**Figure S10**).

The electrocatalysts were then drop-casted on carbon paper to screen the electrochemical performance. The overpotential of the electrocatalysts was measured in a three-electrode cell using 0.5M $H_2SO_4$ (**Figure 4b, Table S3**). The experimental overpotentials matched well with the trend of predicted activity (**Figure 3c**, **SI note 3**). We observe a coefficient of determination of 0.71 between our theoretically derived quantum mechanical proxy and in-lab experimental overpotential of catalyst candidates (*p-value* < 0.001). We use the structures as obtained by optimization of (9) to model every chemical composition. Furthermore, to compare the intrinsic activity of the electrocatalysts, we decouple the morphology effect from the activity by normalizing the current density by the active electrochemical surface area (ECSA) overpotential instead of the geometric area (**Figure 4c, Figure S11**). We then calculate the mass activity of the electrocatalysts normalized by the Ru amount in the electrocatalyst (**Figure 4d**). Taking into consideration both measures, we identified $Ru_{0.58}Cr_{0.25}Mn_{0.09}Sb_{0.08}O_2$ as the most promising candidate with the highest mass activity 381 A/$g_{Ru}$ (*ca.* 8 times higher than $RuO_2$) and a much lower overpotential increase rate of 2 mV.hr$^{-1}$, 10 times slower than $RuO_2$ (**inset in Figure 4g, Table S3**).[63] To better assess the stability of the electrocatalyst, we prepared it by spraying the electrocatalyst on carbon paper instead of drop-casting, the technique we used for screening purposes, to ensure a higher penetration of the particles through the hydrophilic carbon paper yielding in higher surface coverage and better mechanical attachment to the carbon fibers. We maintained the loading of the electrocatalyst between both preparation techniques at 1 mg/cm$^2$. Our candidate maintained an overpotential of less than 300 mV for 180 hours at a low overpotential increase rate of 2 mV/hr, 10 times lower than $RuO_2$ and any previous reports of rutile structure based catalysts[43]. Finally, we also search existing databases (OQMD, AFLOW, Materials Project, OCP, ICSD) and literature for similar alloys and we were unable to find even a similar oxide compound with Ru, Cr, Mn and Sb present together. Our QCE method was therefore able to effectively search chemical space, identify a new promising family of multi-metal oxides for exploration and discover a highly efficient and stable OER catalyst.

We also compare the results to a random sampling of chemical composition strategy. We observe that 5/8 of our predictions demonstrate an overpotential of less than 300 mV indicating and 3 of them demonstrated an exceptional stability of less than 10 mV/hr overpotential degradation (**SI table 3**). If we had to randomly sample the chemical space to find at least one of these 3 compositions, we would need to perform at least ~766 experiments in the quaternary chemical space of just a single chemical subspace (**SI note 8** for stoichiometric space size); indicating at least x95 improvement over a random chemical space search strategy purely based on experiments.

**Origins of the catalyst stability and efficiency:**

To better understand the cause of higher activity and stability of our top catalyst candidate: $Ru_{0.58}Cr_{0.25}Mn_{0.09}Sb_{0.08}O_2$ (henceforth referred to as catalyst $RuCrSbMnO_2$), we performed detailed DFT calculations and analysis. We used the crystal structure for each candidate, as obtained by minimizing the joint Hamiltonian in (9), for performing the following analysis (**Figure S12-S13** for crystal structure of $RuCrMnSbO_2$ and its XRD comparison to experimental structure). Bader charge analysis showed that the average partial charge on Ru increased from **1.73|e|** in pure $RuO_2$ to **1.82|e|** in $RuCrMnSbO_2$. Ru with a higher partial charge indicates improved ability of the catalyst to oxidize water to oxygen i.e., OER. Additionally, to explore the origin of the stability of the catalyst candidates, we further analyzed the density of states (DOS). As shown in **Figure 5 (a, b)** incorporation of Cr, Mn and Sb alter the DOS of $RuO_2$, causing occupation at Fermi level to decrease from **38 states/spin·cell** to **18 states/spin·cell**. This decrease in DOS at fermi energy indicates stronger bonding between metal atoms, leading to stabilization of our solid solution[43,65] (**Figure S14** for a comparison study that considers only Cr and Mn). **Table 1** summarizes our observations against previously reported $Ru_{0.4}Cr_{0.6}O_2$ and $RuO_2$ [43]. We also observe that entropic contributions further stabilize the predicted composition $RuCrMnSbO_2$ as compared to previously reported $RuO_2$ and $Ru_{0.4}Cr_{0.6}O_2$. These observations align with our observed experimental measurements of 180 hours stability for $RuCrMnSbO_2$ as compared to unstable $RuCrO_2$ and 40 hours stability of pure $RuO_2$ (catalyst ID *R8-M16-16* and *RuO_2* for detailed experimental results on stability and activity in SI tables 1-3).

| Compound | $\Delta H_{mix}$ (eV/atom) | Bader charge on Ru | DOS at $E_{fermi}$ (states/spin·cell) | Entropy contribution at 550° C (eV/atom) |
|---|---|---|---|---|
| $RuCrMnSbO_2$ | -0.08 | 1.82 | 18 | -0.078 |
| $RuCrO_2$ | -0.05 | 1.92 | 19 | -0.048 |
| $RuO_2$ | 0.0 | 1.73 | 38 | 0.000 |

**Table 1**: Comparison of different stability contributing factors for $RuO_2$, $Ru_{0.4}Cr_{0.6}O_2$ and $RuCrMnSbO_2$

At the same time, we calculated free energy profiles ($\Delta G$) of OER to compare the activities of $RuCrMnSbO_2$ to $RuO_2$ (refer to Data Generation in Methods for details and **SI note 5-6**). As can be seen from **Figure 5 (c,d)**, the formation of OOH was found to be the rate-determining step. For $RuCrMnSbO_2$, the bader charge was found to be 1.92 eV which is smaller than the bader charge corresponding to $RuO_2$ (2.02 eV) and consistent with the experimental observation of smaller overpotentials. This observation can be explained by the finding that the d-band surface density of states (DOS) is higher for $RuCrMnSbO_2$ (**7.52 states/spin·cell**) as compared to $RuO_2$ (**5.30 states/spin·cell**) leading to the lower overpotential and lower reaction energy barrier to the formation of adsorbed OOH radical. Presence of larger number of states enable easier electron transfer and therefore, facilitates formation of intermediates (*O, *OH, *OOH)[64].

**Conclusion:**
In this study, we reported a new approach for materials discovery that maps the chemical space search problem to one of finding the ground state of the Ising Hamiltonian. This enables us to use a quantum-inspired computing framework to find materials with optimal properties in an accelerated and efficient fashion, orders of magnitude faster than the wide used alternatives such as genetic algorithms and Bayesian optimization. Our efforts led us to develop an improved efficiency proxy and discover a stable and efficient OER catalyst **$Ru_{0.58}Cr_{0.25}Mn_{0.09}Sb_{0.08}O_2$**. Post-hoc DFT analysis further explains the electronic origins of the stability and efficiency of the catalysts.

However, we should also take note of one of the limitations of the current study: restriction of one lattice across the chemical space, which is not what is always desired in materials discovery pipelines. Nevertheless, this can be handled by using extended cluster expansion mapping as proposed by Koretaka Yuge[66]. We plan to incorporate this as part of our future work.

# Methods:
**DA:**
Fujitsu's Digital Annealer (DA) is designed to efficiently solve combinatorial optimization problems[27] formulated as fully connected Ising problems expressed in Quadratic Unconstrained Binary Optimization

(QUBO) form. In most general case, a QUBO can be represented as: $F(q_0, q_1, \ldots, q_n) = \sum_{ij} a_{ij} q_i q_j$ where $q_i$ $(i = 1 \ldots n)$ are binary variables.

In this work, we used third-generation DA that can handle up to 100k binary decision variables. Given a QUBO expression, the DA finds the assignment of binary variables (referred by $q_0, q_1, \ldots$) such that the expression value is minimized. All experiments were conducted on the DA environment prepared for research use. More details on exact algorithms and configuration of DA can be found at: https://www.fujitsu.com/jp/documents/digitalannealer/researcharticles/DA_WP_EN_20210922.pdf

**Data generation:**

We use $RuO_2$ (Materials Project[67] ID: mp-825) and $ZrO_2$ (Materials Project ID: mp-2858) as the prototype structures and generated 72 and 96 atoms supercells respectively for training the cluster expansion models. 120 randomly substituted alloy structures were generated for each of the chemical spaces spanned by Ru-Ti-W-Sb-Cr-Mn-V-Co and Ru-Zr-Hf-Y-V-Co-Fe-Ce. We further randomly selected 34 structures and generated surfaces with 110 orientations for the calculation of adsorption energies with vacuum spacing of 10Å on both sides. We chose 110-orientation since we found it to be most stable with respect to surface energy (refer to **SI note 6**). The slab models were composed of 144 atoms (48 metal atoms and 96 oxygen atoms) repeated in 4 layers besides the adsorbates (O, OH and OOH). DFT calculations were performed in VASP for the bulk and surface structures using Perdew-Burke-Ernzerhof[68] (PBE) exchange-correlation functional augmented with Hubbard Coulomb interaction potential (U) corrections for d-electrons taken from Materials Project[67]. Valance electrons were described with a 520-eV plane-wave basis set and 0.05 eV Gaussian smearing of the electronic density. All the bulk geometries were optimized with the energy convergence criterion of $10^{-4}$ eV and force convergence of 0.03 eV/Å. Core electrons were described using the projector augmented wave (PAW) method. All the calculations were spin-polarized and reciprocal space was simulated using 3x2x1 k-point mesh centered at gamma. Refer to **SI note 2** for a detailed evaluation of the computational costs associated with a single calculation.

For Cu-Pd-Ni-Ag chemical space benchmark, we used Cu-FCC 64 atoms supercell as the prototype structure. 120 randomly decorated structures were generated. DFT calculations were performed in VASP for the bulk using Perdew-Burke-Ernzerhof[68] (PBE) exchange-correlation functional. Valance electrons were described with a 520-eV plane-wave basis set and 0.05 eV Gaussian smearing of the electronic density. All the geometries were optimized with the energy convergence criterion of $10^{-4}$ eV and force convergence of $10^{-3}$ eV/A. Core electrons were described using the projector augmented wave (PAW) method.

To perform post-hoc DFT analysis of the $RuO_2$ and $RuCrMnSbO_2$, we generate surfaces with orientations: [001, 100, 101, 110] and choose 110 as the orientation to perform adsorption energy calculations due to lower energy (SI note 5). The post-hoc DFT analysis was done on a k-points grid 3×2×1 with 520 eV as the energy cut-off for the PAW pseudopotentials and SCAN exchange-correlation functional was used with the Hubbard-U correction. The rest of the parameters stayed the same as in the preceding paragraph.

**Mixing energy ($\Delta H_{mix}$):**

We define mixing energy or enthalpy of mixing $\Delta H_{mix}$ as excess energy with respect to the single metal precursors. For metal alloys it assumes the form:

$$\Delta H_{mix}(A_x B_y C_z D_w) = E(A_x B_y C_z D_w) - \frac{x \cdot E(A_u) + y \cdot E(B_u) + z \cdot E(C_u) + w \cdot E(D_w)}{u}$$

where $u = x + y + z + w$ and $E(\cdot)$ represents total energy of the concerned phase.

Similarly, for mixed metal oxides, it assumes the form:

$$\Delta H_{mix}(A_xB_yC_zD_wO_v) = E(A_xB_yC_zD_wO_v) - \frac{x \cdot E(A_uO_v) + y \cdot E(B_uO_v) + z \cdot E(C_uO_v) + w \cdot E(D_wO_v)}{u}$$

where $u = x + y + z + w$.

**Cluster expansion training:**
We used ICET to train cluster expansion models[37,69]. The scripts and codes are available at https://github.com/hitarth64/quantum-inspired-cluster-expansion. We tested different cluster generation schemes by tuning the kind of interactions and cut-off for each of the interactions (**Figure S5** for a comparison plot). Parameters that lead to lowest cross-validation error were chosen for analysis.

**Encoding schemes for QCE:**
Several candidate functions can be used for the encoding scheme $C(\cdot)$. Two natural choices are:
- Encoding 1:

$$\Phi_i(\sigma_i) = \sum_{j=1}^{n} q_j \cdot \phi_i(\sigma_i = M_j)$$

  where $q_j$ are binary decision variables and the one-hot constraint of equation (3) is then enforced as a penalty:

$$P_i(\sigma_i) = \left(\sum_{i=1}^{n} q_i - 1\right)^2$$

- Encoding 2:

$$\Phi_i(\sigma_i) = \sum_{j=1}^{n} f_j(q_1, q_2, \ldots q_k) \cdot \phi_i(\sigma_i = M_j)$$

where $q_j$ are binary decision variables, $k=\text{ceil}(\log_2 n)$ where n is total number of element types. Functions $f_j(\cdot)$'s are defined such that only one $f_j(\cdot)$ is non-zero and others are zero for any $\vec{q} \in \{0,1\}^k$ and each $f_j(\cdot)$ is one at leas for one $\vec{q} \in \{0,1\}^k$. Here are the steps to construct such $f_j$. First, for each possible bit string in the space $\{0,1\}^k$, construct a polynomial that takes one only for the bit string and zero for other bit strings. Next, determine a surjection from the bit strings to element indices {1….n}. Finally, define $f_j$ as the sum of the polynomials correspond to bit strings that mapped to the element j in the previously defined surjection. Such an encoding scheme trivially enforces the one-hot constraint of equation (3) and therefore, we do not need to add it as a penalty. For a 3 elements system example, the bit strings and the corresponding polynomials are {0,0} and $(1 - q_1)(1 - q_2)$, {0,1} and $(1 - q_1)q_2$, {1,0} and $q_1(1 - q_2)$, and {1,1} and $q_1q_2$. An instance of surjection is $\{0,0\} \mapsto 1$, $\{0,1\} \mapsto 2, \{1,0\} \mapsto 3, \{1,1\} \mapsto 3$. In this case, $f_1 = (1 - q_1)(1 - q_2)$, $f_2 = (1 - q_1)q_2$, $f_3 = q_1(1 - q_2) + q_1q_2 = q_1$.

**Performance benchmark algorithms**:
We perform comparison against 4 heuristic search algorithms that are widely used in chemistry and material science. Their implementation details are as follows:
- **GA:** Genetic algorithm is implemented using DEAP framework[45]. We use eaSimple algorithm with 300 generations and a population size of 300 to run and perform the searches. We use crossover and mutation probabilities of 0.3 and 0.7.
- **GAMuPlusLambda:** This search method uses $(\mu + \lambda)$ evolutionary algorithm to perform the chemical space search using eaMuPlusLambda implementation as provided in DEAP. We use 300 generations and a population size of 300 with crossover and mutation probabilities same as above. Corresponding $\mu$ and $\lambda$ parameters that yielded the most optimal solution were found to be 300 and 400.

- **Bayesian Optimization:** We conduct bayesian optimization based searches to search and optimize the alloy combinations using scikit-optimize[46]. We specifically used Gaussian Processes combined with a hybrid acquisition function that randomly chooses either LCB, EI or PI at every iteration.
- **Gurobi:** We used commercial Gurobi 9.5.2 solver[70] and solved the chemical space optimization as a quadratic integer program. Gurobi is available free of cost for academic purpose.

The exact values are tabulated below:

| Method | Global Optimal solution (eV/atom) | Ground state | Time (secs) | Generation | Population size | BO iterations | Iteration of saturation | Params |
|---|---|---|---|---|---|---|---|---|
| eaSimple | -0.181 | Cu0.69Pd0.31 | 604 | 300 | 300 | | 300 | |
| DA | -0.200432 | Cu0.5Pd0.5 | 80 | | | | | |
| eaMuPlusLambda | -0.169 | Cu0.65Pd0.35 | 1021 | 300 | 300 | | 277 | mu=300, lambda_=40 0 |
| BayesianOptimization | -0.1421 | Cu0.7Ni0.04Pd0.26 | 3600 | | | 528 | 222 | |
| Gurobi | -0.0742 | | 20774 | | | | | |

**Chemical subspace search**:

For performing QCE, we know from equation (3) that we rewrite the cluster expansion formulation as,

$$\Phi_i(\sigma_i) = \sum_{j=1}^{n} C(M_j) \cdot \phi_i(\sigma_i = M_j)$$

such that $M_j, j = 1 \ldots n$ represent the set of elements being considered for the site $i$ and

$$C(M_j) \in \{0,1\} \,\forall\, j$$

$$\sum_{j=1}^{n} C(M_j) = 1$$

In order to restrict the sampling to a set of elements, we limit the summation to set of elements of interest to us. If one wants to avoid element $M_s$, one can rewrite the summation as,

$$\Phi_i(\sigma_i) = \sum_{j=1,\ j \neq s}^{n} C(M_j) \cdot \phi_i(\sigma_i = M_j)$$

Such that

$$C(M_j) \in \{0,1\} \,\forall\, j$$

$$\sum_{j=1}^{n} C(M_j) = 1$$

**OER reaction mechanism**:

OER in acidic and neutral medium is a 4-step reaction with the following reaction mechanism[43,50,51]:

$$* + H_2O \leftrightarrow OH^* + (H^* + e^{-1})$$
$$OH^* \leftrightarrow O^* + (H^+ + e^{-1})$$
$$O^* + H_2O \leftrightarrow OOH^* + (H^+ + e^{-1})$$
$$OOH^* \leftrightarrow * + O_2 + (H^* + e^{-1})$$

where $*$ denotes the species adsorbed at the catalyst surface.
The theoretical overpotential ($\eta_{OER}$) is thus defined as,

$$\eta_{OER} = \max\frac{(\Delta G_7, \Delta G_8, \Delta G_9, \Delta G_{10})}{e} - 1.23\ V$$

$$= \frac{max(\Delta G_{OH^*}, \Delta G_{O^*} - \Delta G_{OH^*}, \Delta G_{OOH^*} - \Delta G_{O^*}, 4.92 - \Delta G_{OOH^*})}{e} - 1.23\ V$$

Calculation of $\eta_{OER}$, thus, requires values of adsorption energies for reaction intermediates $O, OH$ and $OOH$. These DFT energy calculations are computationally expensive. Usually, scaling relationships are used to reduce the number of calculations needed[47]. However, scaling relationships break down in the case of alloys. Therefore, here, we perform explicit DFT calculations for all 3 adsorbate structures: *O, *OH and *OOH along with bare surface to calculate the adsorption energies. These adsorption energies along with zero point energies and entropic corrections are then used to calculate $\Delta G_{OH^*}$, $\Delta G_{O^*}$ and $\Delta G_{OOH^*}$ as,

$$\Delta G_{O^*} = E_{DFT}(O^*) - E_{DFT}(*) - E_{DFT}(O) - E_{ZPE}(O) + T\Delta S(O)$$
$$\Delta G_{OH^*} = E_{DFT}(OH^*) - E_{DFT}(*) - E_{DFT}(OH) - E_{ZPE}(OH) + T\Delta S(OH)$$
$$\Delta G_{OOH^*} = E_{DFT}(OOH^*) - E_{DFT}(*) - E_{DFT}(OOH) - E_{ZPE}(OOH) + T\Delta S(OOH)$$

Furthermore, to overcome the limitation of traditional GGA-based calculations for $O_2$ molecule, its free energy is calculated as,

$$G(O_2) = 4.92 + 2G(H_2O) - 2G(H_2)\ (in\ eV)$$

**Calculation of band centers:**

D-band model relates the surface reactivity with the shifts in the d-band center of the catalyst where the d-band center is a single state with energy $\epsilon_d$ that approximates the interaction of participating bands of d-states. In particular, we calculated d-band centers and p-band centers using the first moments as,

$$\epsilon_{M-d} = \frac{\int_{-\infty}^{\infty} ED_{M-d}(E - E_F)\ dE}{\int_{-\infty}^{\infty} D_{M-d}(E - E_F)\ dE} \quad (8)$$

$$\epsilon_{M-p} = \frac{\int_{-\infty}^{\infty} ED_{M-p}(E - E_F)\ dE}{\int_{-\infty}^{\infty} D_{M-p}(E - E_F)\ dE} \quad (9)$$

where M is the corresponding element and $D_{M-y}$ represents partial DOS of orbital y of element M in the system.

**Logistic Regression Analysis:**

We used logistic regression, as implemented in scikit-learn, to investigate the relationship between band centers and overpotentials of the materials. We first calculated theoretical overpotentials (Methods subsection *OER reaction mechanism*) of 34 randomly chosen candidate materials from the two chemical families under consideration (Methods subsection *Data generation*). We train the classifier to distinguish the differential improvement between two catalysts i.e., trained to predict if $\eta_i < \eta_j$ (overpotential of material $i$ is less than that of material $j$). In addition, refer to **Figure S3** to see the variation in intermediate adsorption energies as a function of band centers.

**Materials:**

Metal precursors including ruthenium chloride hydrate, (RuCl3·xH2O), chromium (iii) chloride anhydrous, 99.99% trace metals basis (CrCl2), manganese chloride (MnCl2), tungsten (VI) chloride (WCl6), titanium diisopropoxide bis(acetylacetonate), 75wt. % in isopropanol, Vanadium(iii) chloride (VCl3), and Propylene oxide (≥99.5% GC) were all purchased from Sigma-Aldrich. Nafion® (5 wt% in a mixture of lower aliphatic alcohols and water) and AvCarb MGL190 were used for electrode preparation. All chemicals were used without any further purification.

**Experimental synthesis**:

The mixed metal oxides were synthesized using a modified sol-gel procedure [71] by completely dissolving a total amount of 2.7 mmol of the metal precursor mixture in 3.5 mL anhydrous ethanol. The solution was vortexed for 5 mins and sonicated in a water bath for 1 hour until it is clear. Then, it was chilled in a refrigerator for 2 hours to prevent any undesired hydrolysis and condensation which may affect the gelation process. Afterward, a magnetic stirrer was used to mix the solution vigorously while 2 mL of propylene oxide was added dropwise to the mixture. The solution was aged for 1 day to promote gel formation and then washed with acetone; the process was repeated for 5 days before drying the gel in a vacuum oven for 1 day. The sol-gel process produces amorphous oxides, therefore, to produce a rutile crystal structure the dried powder was annealed at 400°C for 2 hours.

For the fast electrochemical purpose, the working electrode was prepared on 0.5 cm x 0.5 cm untreated carbon paper (AvCarb MGL190) by drop casting. First, a catalyst ink was prepared by mixing 10 mg of the catalyst in 1 mL mixture of water and ethanol (4:1, v/v). Then, the ink was sonicated in an ice bath for 1 hour. Finally, a 25 μL drop of the ink was deposited on the carbon paper and allowed to dry in the air. For final electrochemical testing, the working electrode was prepared by spraying the electrocatalyst using $N_2$ on 0.5 cm x 0.5 cm untreated carbon paper. The ink was prepared by mixing 10 mg of the catalyst in 1 mL of isopropanol. Then, the ink solution was sonicated in ice bath for 1 hour before spraying it on the carbon paper on the hot plate at 90 °C.

**Electrochemical testing**
Electrochemical testing was conducted in a three-electrode setup using a $Hg/Hg_2SO_4$ as the reference electrode and graphite rod as the counter electrode. The OER activity was evaluated by running linear sweep voltammetry (LSV) with a rate of 5 mV/s. The stability of the catalyst was evaluated by conducting chronopotentiometry at 10 mA/cm².

All potentials were iR-corrected by measuring the solution resistance from electrochemical impedance spectroscopy (EIS) with a bias of 1.53V vs. RHE in the frequency range from 100 kHz to 10 MHz and an amplitude of 5 mV. All the potentials in this study were reported with respect to a reversible hydrogen electrode (RHE) using the following relationship:

$$E_{RHE} = E_{Hg/Hg_2SO_4} + 0.640 + 0.0591 \times pH$$

**Structure analysis**
The crystal structure of catalysts was determined using X-ray Diffraction (XRD). A Miniflex 600 (Rigaku, Japan) equipped with D/tex Ultra silicon trip detector and Cu Kα radiation (λ = 1.5418 Å) was used. Powders were prepared by mixing with acetone and then dropping a small drop of the mixture to fill a 4 mm diameter x 100 μm deep groove in a single crystal silicon holder (zero-background). The angle was varied between 20° to 80° with a step size of 0.005° and a scan rate of 1 deg/min.

To evaluate the nanocrystalline size and macrostrain of the electrocatalysts, we used Scherrer's equation and microstrain equation:

$$D = \frac{K\lambda}{FWHM \cos\theta}, \text{ Scherrer's equation}$$

$$\varepsilon = \frac{FWHM}{4 \tan\theta}, microstrain$$

D is the mean crystallite size, λ is 0.154 nm for Cu X-ray source, K is shape factor has a typical value between 0.9-1. The full-width half-maximum (FWHM) of the peaks were measured by fitting to Gaussian distribution and then calculated from the standard deviation (σ) using the following relation:

$$FWHM = 2\sqrt{2.\ln 2}\ \approx 2.355\,\sigma$$

**Electron microscopy**

The structural characterization and elemental mapping of the catalysts were done using High Resolution Scanning Electron Microscopy (HRSEM) and Transmission Electron Microscope (TEM). The experiments were conducted in a Hitachi HF3300 equipped with a cold field emission electron gun using an accelerating voltage of 300 kV. Energy X-ray dispersive spectroscopy (EDS) detector was used in scanning transmission electron microscopy (STEM) mode to analyze and quantify the composition of the nanoparticles. Also, a secondary electron (SE) detector was used to collect HRSEM images of the nanoparticles. Powder samples were prepared in ethanol and sonicated for 10 minutes before drop casting a 1 -2 µL drop on a 400-mesh copper grid and drying overnight.

**Data and Code Availability:**

All the data supporting this study are available at https://github.com/hitarth64/quantum-inspired-cluster-expansion

All the slab calculation data is available at: https://github.com/hitarth64/MixedMetalOxides

**Figures**:

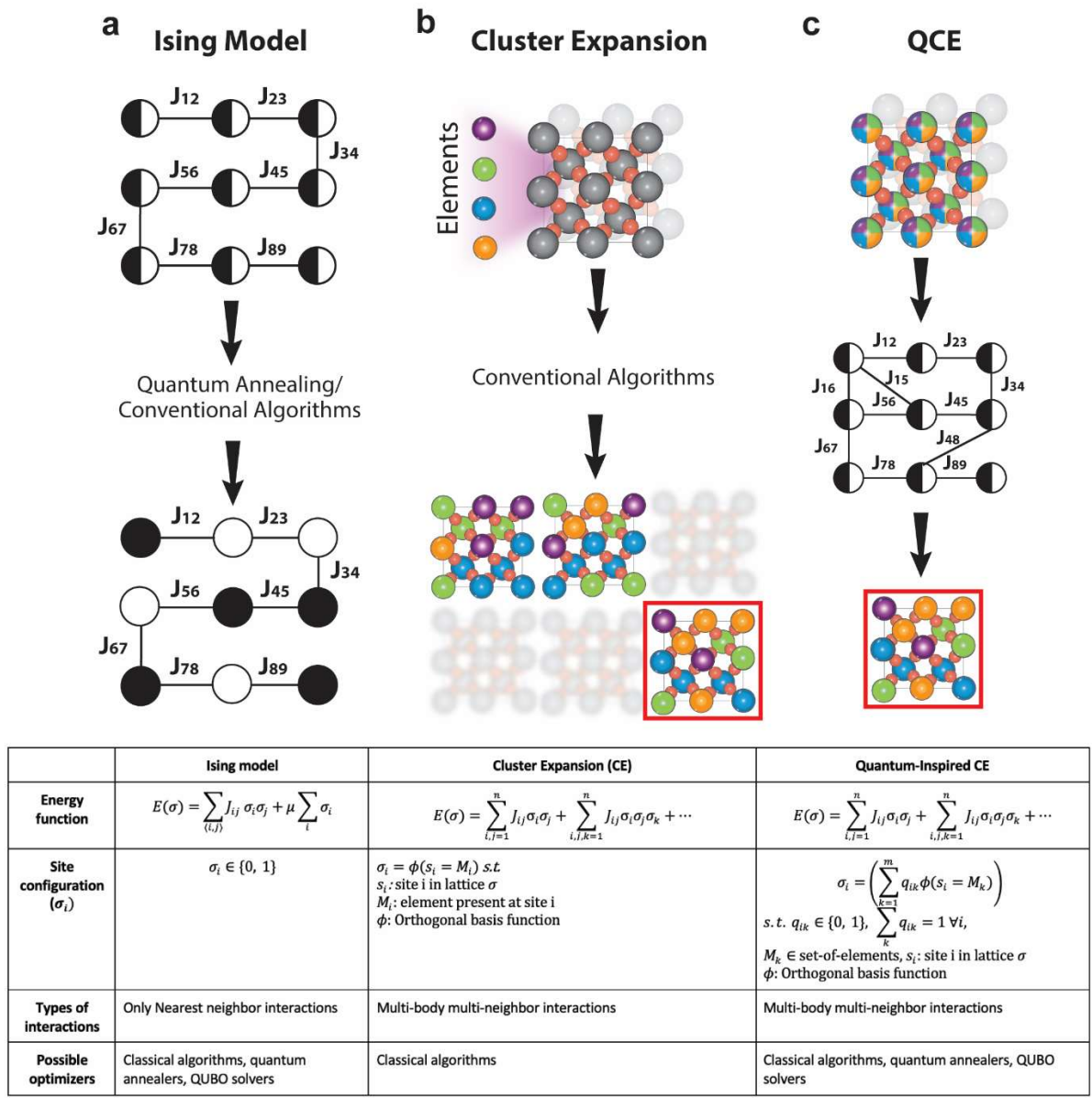

**Figure 1**: **Mapping chemical space search to finding the ground state of an Ising model with Quantum-Inspired Cluster Expansion (QCE)**

Figure shows and compares the Ising model (1st column), cluster expansion (2nd column) and our proposed approach (3rd column). Ising models are computational models of ferroelectricity. Various optimizers including quantum annealers, quantum-inspired optimizers as well as conventional algorithms such as Metropolis can be used to find ground state of Ising models. On the other hand, Cluster expansions (CE) are used for solving the combinatorial search problem in chemical space but only conventional algorithms have been used so far for exploring CE. As part of this paper, we propose a mapping (3rd column of inset table) from CE to an Ising model. This enables us to lever same set of optimizers for CE as those are available for finding ground state of Ising models.

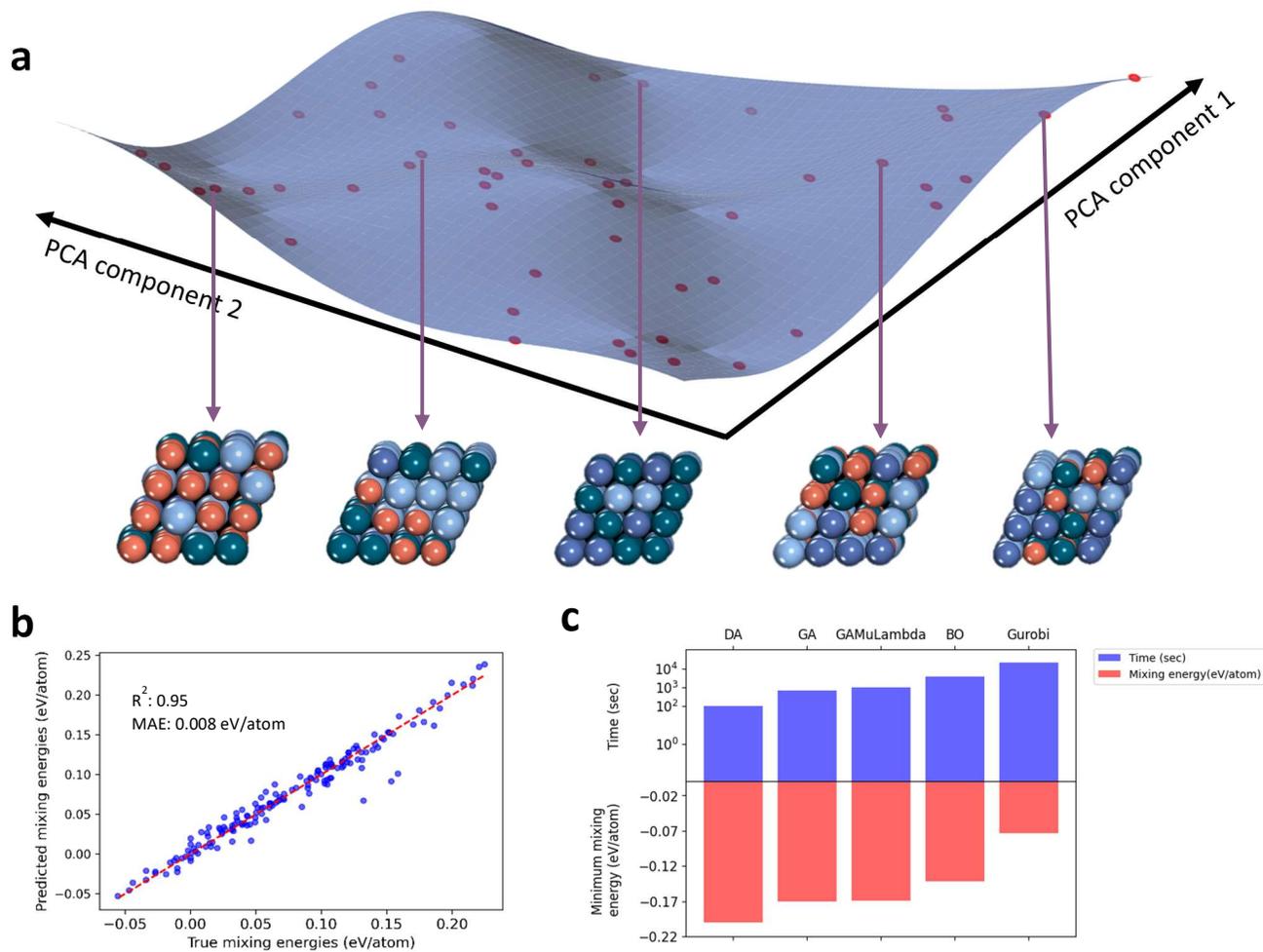

**Figure 2: Chemical space exploration using proposed mapping**

Exploring chemical space with random sampling often ends up with limited exploration. Accurate surrogate models and search algorithms can help us explore unsampled chemical territories. (a) Shows the mixing energy landscape in a 2D projection of the chemical space. Red dots correspond to the sampled points and each of the points in this 2D projection correspond to a chemical structure. As we can see the global minima has not been sampled. (b) Cluster expansion predictive accuracy for quaternary alloy composed of Cu-Ni-Pd-Ag (our benchmark materials system). (c) Performance of our QCE approach using DA against other widely used algorithms. We perform chemical ternary subspace search in Cu-Pd-Ni space. We observe significant temporal acceleration against genetic algorithms (GA) and its variant MuPlusLambda in addition to bayesian optimization (BO) and a commercial optimization solver Gurobi. The ground states found by the different solvers vary with only DA providing the actual ground state configuration and composition. Refer to methods subsection on benchmark algorithms for details on implementation.

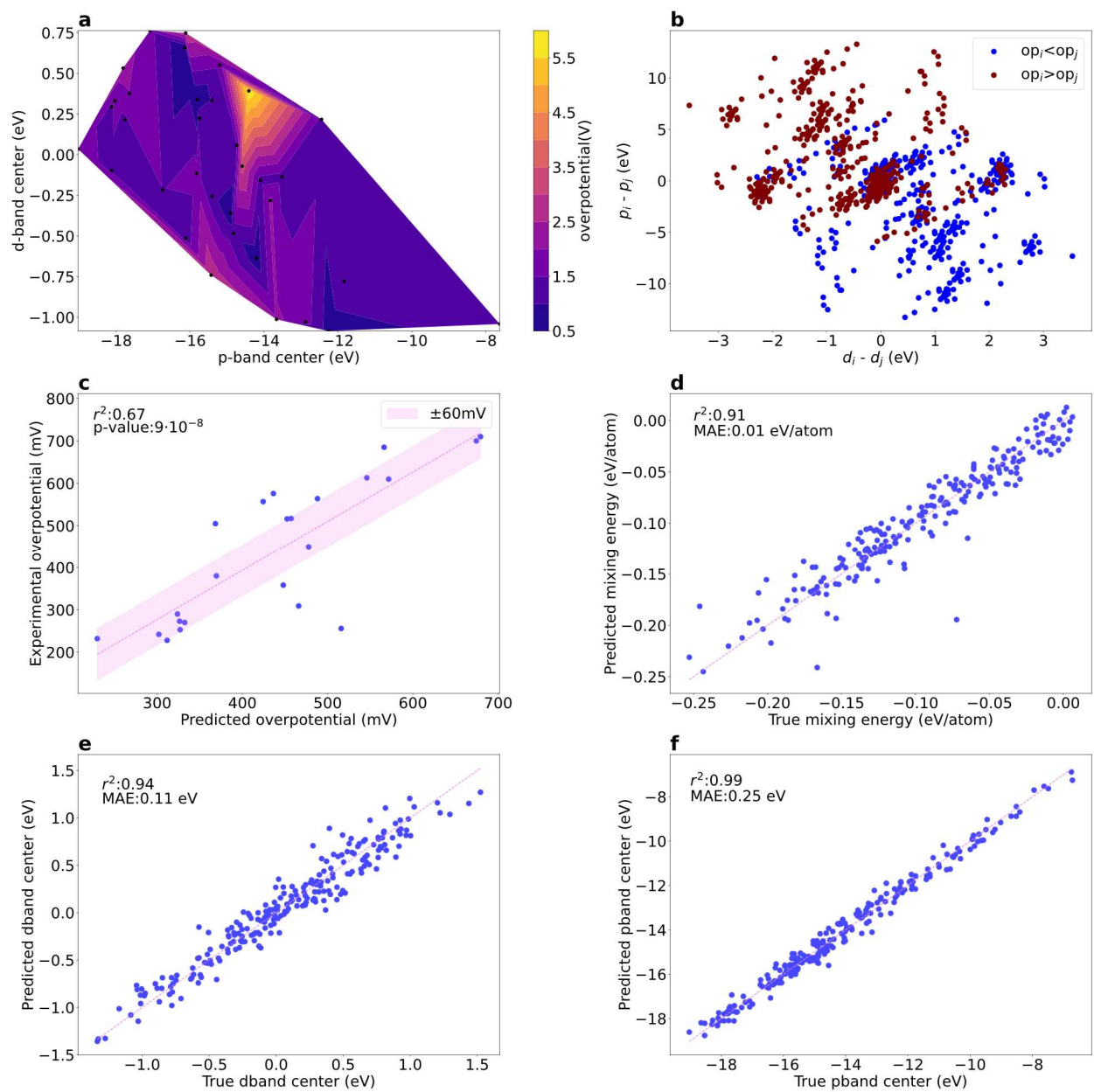

**Figure 3: Development of an efficiency proxy for OER catalysts and training the cluster expansions.**

(a) plot investigates the relationship between electronic structure of the material and its efficiency expressed through overpotential. We trained a logistic regression based classifier to compare if one material has a larger overpotential than the other (0 if smaller and 1 if larger) and use it to search for more efficient materials; (b) We further validate this hypothesis on overpotentials obtained through in-lab experiments as part of this study; when represented as band centers, we are able to cluster the differences with high accuracy; (c) shows how the theoretically calculated d and p band centers correlate with experimental overpotentials. We use the exact coefficients obtained through logistic regression comparison classifier to perform this analysis. (d)-(f) shows the performance of cluster expansion models used for exploration within this study for mixing energies, d&p band centers of the bulk structures for [Ru-Cr-Mn-Sb-Ti-V-W-Co]O$_2$

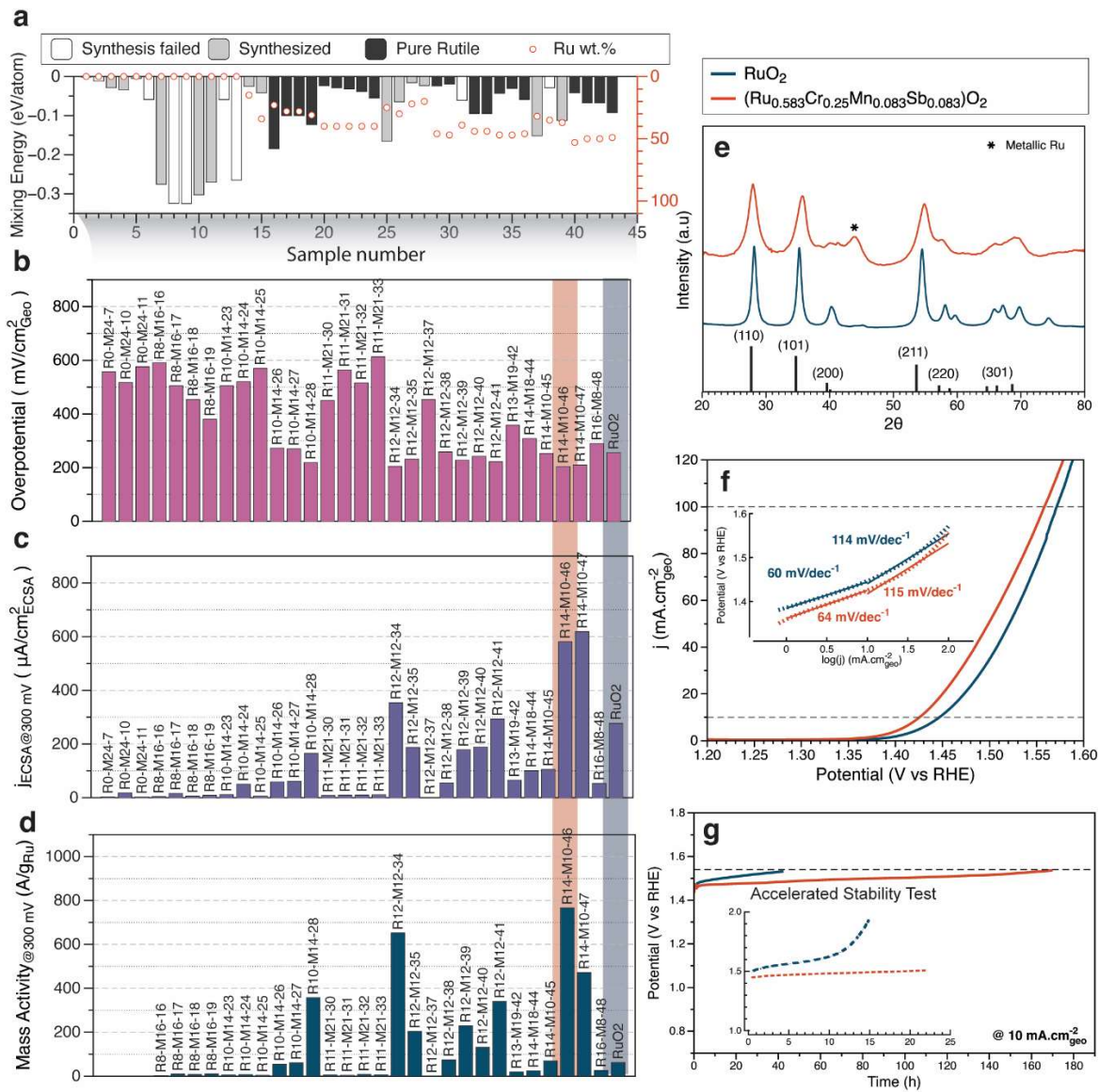

**Figure 4**: **Experimental screening of electrocatalysts.**

(a) The mixing energy of all electrocatalyst candidates considered for experiments in this study. The candidates were classified based on the synthesis outcome: synthesis failed (the electrocatalyst wasn't produced), synthesized, and pure Rutile phase. (b) Overpotential screening of synthesized electrocatalysts. (c) The intrinsic activity of electrocatalysts normalized by the electrochemical surface area. (d) The mass activity of electrocatalysts normalized by the total mass of Ru. (e) XRD and (f) polarization curves (Tafel slope analysis in the inset) of the best electrocatalyst candidate R14-M10-41 ($Ru_{0.58}Cr_{0.25}Mn_{0.083}Sb_{0.083}O_2$) and a baseline $RuO_2$ catalyst. (g) Chronopotentiometry test of the electrocatalysts sprayed on carbon paper electrodes. In the inset is the accelerated test conducted on the electrocatalysts drop-casted on carbon paper. The electrocatalyst loading for all samples is 1 mg/cm². The electrolyte is 0.5 M $H_2SO_4$.

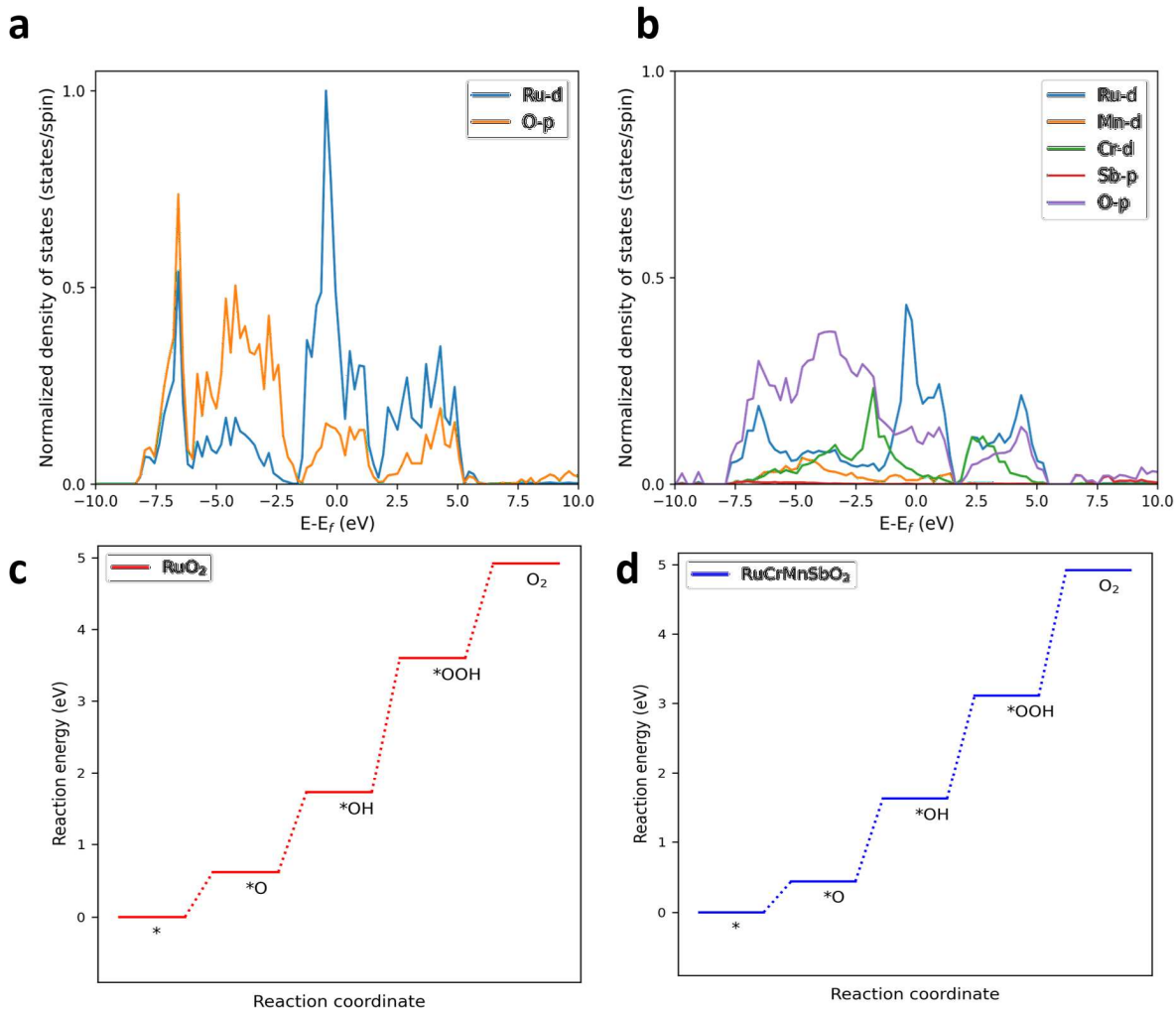

**Fig. 5**: **Post-hoc DFT analysis of the best candidates**

Figures (**a, b**) show bulk DOS decomposed into elemental contributions and normalized to one for all the constituent elements for RuO$_2$ and RuCrMnSbO$_2$. The density of states for RuO$_2$ at fermi level is 38 states/(eV·spin) larger than 18 states/(eV·spin) in C$_1$ which is an indicator of better chemical bonding[65]. Figure (c,d) show free energy diagrams for OER with both the catalysts (RuO$_2$ and RuCrMnSbO$_2$). In both cases, *OOH formation is the Rate Determining Step (RDS) with $\eta_{RuO_2} > \eta_{RuCrMnSbO_2}$ where $\eta$ represents overpotential.

**Acknowledgements:**

This work was financially supported by Fujitsu Ltd. and Fujitsu Consulting (Canada) Inc. We thank Fujitsu for their generous access to the Digital Annealer where all the annealing operations and optimizations were performed. All the DFT calculations were performed on the Compute Canada's Niagara supercomputing cluster. The experimental work was supported financially by the Natural Sciences and Engineering Research Council (NSERC) of Canada, Vanier Canada Graduate Scholarship. Theoretical work was supported through Hatch Scholarship for Sustainable Energy Research. Electron microscopy was conducted at the Ontario Center for the Characterization of Advanced Materials (OCCAM).

**Competing interests:**

The authors declare no competing interests.